\documentclass[11pt,aps,prd,reprint]{revtex4}
\usepackage{epsfig}
\usepackage{graphicx}
\usepackage{amsfonts}
\usepackage{latexsym}
\usepackage{amsmath,amssymb}
\usepackage{verbatim}
\usepackage{mathtools}
\usepackage{setspace}
\usepackage{slashed}
\usepackage[all]{xypic}

\usepackage{tikz}
\usepackage{psfrag}
 \usepackage{subfig}
\usepackage{comment}
\usepackage{array}
\usepackage{amssymb}
\usepackage{amsmath}
\usepackage{amsthm}
\usepackage{graphicx}
\usepackage{float}

\begin{document}
	
\title{A toy model of viscous relativistic geometrically thick disk in Schwarzschild geometry }
\author{Sayantani Lahiri}

\author{Claus L{\" a}mmerzahl}
\affiliation{University of Bremen, Center of Applied Space Technology and Microgravity (ZARM), 28359 Bremen}
\email{sayantani.lahiri@zarm.uni-bremen.de; claus.laemmerzahl@zarm.uni-bremen.de}

\date{\today}

\begin{abstract}
	{\noindent 
\textbf{Abstract} : In this earliest study of thick accretion disks with viscosity effects, we construct stationary solutions of a relativistic geometrically thick accretion disk in the Schwarzschild spacetime under the influence of shear viscosity and the curvature of the black hole by solving the general relativistic causal Navier-Stokes equation.
Motivated by the causal prescription of relativistic hydrodynamics initially introduced in M{\"u}ller-Israel-Stewart theories, our approach adopts a simplistic path and takes into account of only shear viscosity, discarding influences of bulk viscosity and heat flow.
This work investigates possible impacts of both the shear viscosity tensor and black hole curvature on the shape of a thick disk characterized by constant specific angular momentum distribution.  
The existence of the integrability condition of the Navier-Stokes equation has been examined in our study which further supports the existence of stationary solutions in the given set-up. }
\end{abstract}

\maketitle


\section{Introduction}
Till date, the process of matter accretion attributed by transport of astrophysical fluids towards black holes is considered as one of the primary sources of energy release in the universe produced as a result of conversion of the gravitational energy of the in-falling accreting fluid into radiation. 
Accretion disks are presumed to exist in various astrophysical environments like active galactic nuclei, quasars, ADAFs, X-ray binaries and gamma-ray bursts \cite{Frank,Zanotti} . 
The study of an accretion disk can be broadly divided into two categories :  study of equilibrium stationary solutions in a given black hole background and investigation of time dependent processes which involve the dynamics of in-falling matter within the black hole and hence the accretion process. 
Various models of accretion disks have been developed over past years, for example, geometrically thick disks or Polish doughnuts \cite{Kozlowski}, thin disks \cite{Novikov}, slim disks \cite{Abro-1} for acquiring comprehensive knowledge about the black hole accretion process as well as related physical processes occurring within a disk in a given background geometry.      
 In order to determine various properties of accretion, a primarily requirement is the specification of the matter content of the disk which in the simplest case involves the disk being filled with ideal fluid.
More general scenarios correspond to consideration of electric field \cite{Kovar} in the ideal fluid, possible interplay of magnetic fields \cite{Pugliese,Soler}, viscous and turbulent processes and radiation processes.
The study of an accretion disk also involves adoption of different approaches namely the test particle approach, the kinetic description for describing magnetic plasma, $\alpha$-viscosity prescription, all depending on particular regimes and time scales under consideration. 
Among various disk models proposed so far, the relativistic geometrically thick disk modeled by an ideal fluid, with negligible self gravity, is one of the simplest analytically studied stationary equilibrium model of in a black hole background\cite{Kozlowski}.
In the traditional approach, any role of dissipative processes are ignored.\\
In the present work, we aim to construct stationary solutions of geometrically thick accretion disk described by constant angular momentum distribution in presence of viscosity around a Schwarschild black hole.
The self gravity of the disk is neglected throughout the study. Assuming that a single species of particle exists in the fluid, the disk is modeled with a non-ideal fluid involving shear viscosity. All hydrodynamical equations are expressed in the Eckart frame, a common choice of frame in astrophysical contexts. 
We have deliberately imposed minimalistic modifications due to viscous effects and introduced the shear viscosity as perturbation to the ideal fluid configuration. 
The condition of stationarity is therefore not violated in presence of viscosity, as a result, the fluid is assumed to rotate in circular orbits. The bulk viscosity and the heat flow within the non-ideal fluid are neglected for simplicity. 
The primary aim of this work is to study the role of shear viscosity in governing the shape of thick disks. 
Moreover, the novelty of our approach is that the general form of shear viscosity tensor is endowed with curvature of the central black hole, as a result, the
shape of a relativistic thick disk will also be governed by the curvature and hence will be different for various black hole geometries. 
So the propereties of the thick disk will be straight-way influenced by the curvature of the particular black hole geometry in which the disk resides.   \\ 
It is already known that the relativistic angular momentum conservation equation of non-ideal fluids, i,e. the Navier-Stokes equation does not preserve causality. 
This problem is circumvented with the help of the formulation developed by M{\"u}ller and corresponding relativistic extension by Israel-Stewart \cite{Isr-1, Isr-2}, known as causal relativistic hydrodynamics. It is a phenomenological approach that introduces additional transport coefficients and resulting conservation laws are stable.\\ 
In the present work, we will consider causal theory of relativistic hydrodynamics by taking into account of curvature, first formulated in the context of relativistic heavy ion collision \cite{Romatschke-1,Baier}. 
Very recently, the the causal theory of relativistic viscous hydrodynamics is extended in the Eckart frame using gradient expansion scheme \cite{Lahiri} in which general forms of dissipative flux quantities have been constructed upto second order in gradients of hydrodynamical variables such that,
as one of the second order gradient terms, the curvature terms appear in each of these quantities. \\
There are several advantages of our approach we adopt here. 
First, the curvature of  black hole has a direct impact on its shape thereby facilitating the investigation for searching effects of viscosity as well as curvature on a geometrically thick disk. 
Thus our work gives an opportunity to compare our results with the ideal fluid scenarios. 
Second, viscous effects on a thick disk are not independent of the central black hole but varies according to black hole geometry in which the disk is located. 
Third, as the curvature explicitly enters in the equations of motion, a probe for strong gravity effects on the thick accretion might also be possible.\\ 
Although the set-up examined here does not provide the complete picture of viscous effects playing within a disk since we avoided turbulence, magnetic fields, effects of heat flow, contributions arising due to non-linear terms of shear tensor, expansion scalar, vorticity and alike,
nevertheless, our work can be conceived as a toy model which would help us visualize quantitative influences of shear viscosity (in its linear order) on the shape of the thick disk in presence of a constant specific angular momentum distribution of the disk.\\ 
The work is categorized as follows. In section II, we briefly discuss the construction of the general form of viscosity tensor that gives rise to causal Navier-Stokes equation in the Eckart frame. In section III, we constructed stationary thick disk by solving general relativistic causal Navier-Stokes equation under the influence of spacetime curvature and constant angular momentum distribution. 
The validity of the integrability condition of the causal Navier-Stokes equation is also examined.  Finally a conclusion and possible outlooks are presented in section-IV.  
\section{Description of geometrically thick disk in presence of curvature and shear viscosity : Set-up }
Before plunging ourselves into in-depth discussions about non-ideal fluids for describing viscous effects arising within relativistic thick disks, let us briefly recall relevant aspects of stationary, relativistic thick disk in presence of ideal fluid \cite{Kozlowski,Abro-1} obeying a barotropic equation of state to enable self-consistency of our work.
It is assumed there exists a single species of conserved charge so that the particle four-current and the energy momentum tensor of the ideal fluid are given by,
\begin{eqnarray}
N_{ideal}^{\mu} = n u^{\mu}, \qquad
T_{ideal}^{\mu \nu} =  e \, u^{\mu} u^{\nu} + p \,\triangle ^{\mu \nu} \label{current} 
\end{eqnarray} 
where $e$, $p$, $n$ are respectively the total energy density, fluid pressure and number density of particles.
The projection tensor is defined as $\triangle ^{\mu \nu}=g^{\mu \nu}+u^{\mu}u^{\nu}$ such that $u_{\mu}\triangle^{\mu \nu}=0$ and the normalization condition is given by $u^{\alpha}u_{\alpha}=-1$.
If the covariant derivative is decomposed as $\nabla^{\mu}=-u^{\mu}D + \mathcal{D}^{\mu}_\bot $ where the longitudinal and transverse parts are respectively $D=u^{\alpha}\nabla_{\alpha}$ and  $\mathcal{D}^{\mu}_\bot= \triangle^{\mu \nu} \nabla_{\nu}$, using (\ref{current}), the conservation laws corresponding to mass, energy and momentum can be expressed as follows, 
\begin{eqnarray}
Dn +n \nabla.u&=&0   \qquad \text{(continuity equation)} \label{4},\\
D e +(e+p)\nabla.u &=&0,  \label{5}\\
Du^{\mu}& =&-\displaystyle \frac{1}{(e+p)}\mathcal{D}^{\mu}_\bot p \qquad \text{(Euler equation)}\label{6}
\end{eqnarray}
where the four-acceleration of the fluid is defined to be $a^{\mu}=Du^{\mu}$. 
The stationary relativistic thick disks (tori) supported by ideal fluid in equilibrium obeying a barotropic equation of state and describing circular orbits around the Schwarzschild black hole can be characterized by its angular velocity $\Omega(r,\theta)$ and by a specific angular momentum distribution $l(r,\theta)$. 
Then the dynamics of the fluid flow is governed by the momentum conservation equation which can be expressed in the integral form as follows, 
\begin{eqnarray}
W(r,\theta)-W_{in}(r,\theta)=-\int_{0}^{p}\frac{dp}{e+p}=\ln|u_{t}|-\ln|u_{t}|_{in}-\int_{l_{in}}^{l}\frac{\Omega dl}{1-\Omega l}
\end{eqnarray}
 where "in" corresponds to the inner edge of the disc in the equatorial plane and $W(r,\theta)$ is the total potential in the Newtonian limit consisting of gravitational and centrifugal potentials \cite{Font}. At infinity, $W(r,\theta)=0$. The integration constant is set as $W_{in}=\ln(u_{t})_{in}$. 
 Since we are interested in stationary solutions with constant angular momentum distribution in the equatorial plane, the quantity $W(r,\theta)$ in the Schwarzschild geometry from the above equation reduces to,
 \begin{eqnarray}
 W_{(Sch)}(r,\theta)=\ln (-u_{t})_{(Sch)}=\frac{1}{2}\ln \frac{r^2(r-2M)\sin^2 \theta}{r^3 \sin^2 \theta-(r-2M)l^2(r,\theta)} \label{W-sch}
 \end{eqnarray} 
It is known that the Schwarschild spacetime permits two killing vectors $\eta^{\mu}=(1,0,0,0)$ and $\xi^{\mu}=(0,0,0,1)$ leading to $g_{\mu \nu}\,_{,t}=g_{\mu \nu}\,_{,\phi}=0$ which imply the fluid pressure and the energy density are functions of $r$, $\theta$ only and the functional dependences remain valid even in non-ideal fluids. 
In case of an ideal fuild, the existence of an integrability condition is deduced from the integral form of Euler equation by satisfying the compatibility condition $\partial_{\theta}\partial_{r}p=\partial_{r}\partial_{\theta}p$ \cite{Zanotti} for a given specific angular momentum distribution
and the barotropic equation of state $p=p(e)$. Then the integrability condition helps defining the equipotential surfaces $W(r,\theta)$ which in turn governs the shape of stationary relativistic thick disks.\\
The scenario changes if viscous effects are assumed to be non-vanishing with the disk. 
The most general form of energy momentum tensor and the particle current of a non-ideal fluid in the Eckart frame are given by, 
\begin{eqnarray}
N^{\mu}=nu^{\mu}, \qquad
T^{\mu \nu}=e u^{\mu} u^{\nu}+ (p+\Pi)\triangle^{\mu \nu}+q^{\mu}u^{\nu}+q^{\nu}u^{\mu}+ \pi^{\mu \nu}  \end{eqnarray}  
For simplicity, in the present work, the heat flow $q^{\beta}$ and the bulk viscosity $\Pi$ are neglected so that the Navier-Stokes equation becomes,
\begin{equation}
(e+p)a^{\mu}+\triangle^{\mu \rho} \nabla_{\rho}p+\pi^{\mu \rho}a_{\rho}+\triangle^{\mu}_{\,\beta}\, \triangle_{\sigma \tau} \nabla^{\tau} \pi^{\beta \sigma}= 0
\end{equation} 
where the shear viscosity tensor (traceless part) is given by, 
 \begin{eqnarray}
  \pi^{\mu \nu} =  -2 \eta \sigma^{\mu \nu} = - 2\eta \left(\displaystyle \frac{\mathcal{D}^{\mu}_\bot u^{\nu}+\mathcal{D}^{\nu}_\bot u^{\mu}}{2}-\displaystyle \frac{1}{3} \triangle^{\mu \nu}\nabla.u\right)  \label{visco-1}
  \end{eqnarray}
Here $\eta \geq 0$ is the shear viscosity coefficients.  In the stationary axisymmetric spacetime (Schwarzchild and Kerr) $\eta$ is in principle a function of $r$ and $\theta$ co-ordinates. 
Moveover for circular orbits the expansion scalar vanishes i,e. $\nabla.u =0$.
The relativistic Navier-Stokes equation violate causality due to their parabolic nature allowing superluminal velocity of propagating signals \cite{Romatschke-1}, \cite{Roy} and the associated equilibrium states are plagued with instabilities \cite{Hiscock}. 
M{\" u}ller, later Israel-Stewart(IS) introduced the relativistic version and rectified the causality violating nature of conservation equations of relativistic non-ideal fluids \cite{Isr-1, Isr-2} by considering second order gradients (we note that shear viscosity tensor (\ref{visco-1}) is defined in terms of first order derivatives/gradients in the original formulation by Eckart, this is also true for bulk viscosity where $\Pi=-\zeta \nabla.u$) as a result $\Pi$ and $\pi^{\mu \nu}$ contain additional coefficients arising due to inclusion of second order of gradients. This formulation resulted into hyperbolic equations of motion and causality was thus preserved.  
However the Navier-Stokes equation, being an effective equation can also contain curvature terms in curved spacetime because it is known that Riemann curvature arises due to non-commutivitive property of covariant derivatives as
$
R^{\lambda}_{\:\mu \alpha \beta}\,u_{\lambda} =\nabla_{\alpha}\nabla_{\beta}\,u_{\mu}-\nabla_{\beta}\nabla_{\alpha}u_{\mu}
$.
Thus it is legitimate to consider curvaure terms at the second order gradients in the IS formalism although they were missed out in the original formulation. 
Since we have neglected all effects of bulk viscosity and heat flow, we enlist possible terms involving curvature in the shear viscosity tensor,
\begin{equation}
b_1 R\triangle^{\mu\nu}, \qquad  b_2\sigma^{\gamma<\alpha}\sigma^{\beta>},\qquad \kappa_1 R^{<\alpha \beta>},\quad \kappa_2 u_{\alpha} u_{\beta} R^{\alpha<\rho \sigma> \beta} \label{terms}
\end{equation}
where angular brackets in (\ref{terms}) signify traceless symmetric combinations. For more discussions about the formulation in the Eckart frame, we refer \cite{Lahiri}. Additionally, terms due to couplings between curvature and shear viscosity might also be present but they are discarded as they are third order in gradients. Out of all the terms in (\ref{terms}), the only non-vanishing term in Schwarzschild spacetime involving curvature is $u_{\alpha} u_{\beta} R^{\alpha<\rho \sigma> \beta}$.\\
The goal of this work is to explore effects of the shear viscosity tensor and eventually the influence of spacetime curvature on the structure of geometrically thick disk in the Schwarzchild black hole under the given condition  
where the shear viscosity is introduced in the form of perturbation to the ideal
fluid configuration.
For simplicity in the present study we consider the shear viscosity tensor is built out of curvature terms and 
the causality preserving term i,e. $2\tau_2^{<}D\eta \sigma^{\alpha\beta}\,^{>}$.  All non-linear terms involving shear tensor, vorticity, expansion scalar, etc are neglected. 
also the each of the coefficients are assumed to be constant quantities in the present work which in a more genelized framework may in principle be functions of pressure.
Then the form of shear viscosity tensor is considered as,
\begin{equation}
\pi^{\mu \nu}=\left[-2\eta \sigma^{\mu \nu}-\tau_2 ^{<}D(-2\eta \sigma^{\mu\nu})^{>}+\kappa_2 u_{\alpha} u_{\beta} R^{\alpha <\mu \nu >\beta}\right]
\end{equation}
This suffices our motivation for investigating viscous and strong gravity effects on stationary solutions of relativistic thick disk by considering the causality preserving term and curvature terms.
\section{Thick disk solutions with constant specific angular momentum}
\subsection{Analytical results}	
Let us consider the Schwarzchild black hole described by the following metric,
\begin{equation}
ds^2=-\left(1-\frac{2M}{r}\right)dt^2 +\left(1-\frac{2M}{r}\right)^{-1} dr^2+r^2 d\theta^2+ r^2 \sin^2\theta d\phi^2   \label{metric-1}
\end{equation}
where the geometrical units $G=c=1$ is used throughout the paper.
We are interested in studying stationary solutions of relativistic thick disk with constant specific angular momentum distribution. 
 We propose for simplicity that the shear viscosity only brings about  perturbative corrections to the fluid pressure and does not give rise to radial velocity and other velocity perturbations as a result the fluid takes circular orbits around the central black hole. Then the four velocity of the fluid within the viscous disk has the following form,
\begin{eqnarray}
u^{\mu}= (u^{t},0,0,u^{\phi})  \label{velocity}
\end{eqnarray} 
where $u^{t}$ and $u^{\phi}$ are functions of $r$ and $\theta$ only.
Moreover, the heat flow is also assumed to be small compared to leading order pressure pertutrbations and hence is neglected in the current work.
Restricting ourselves up to the first order in perturbation, the fluid pressure and the energy density are expanded in the following way where $p_{(0)}$ and $e_{(0)}$ are the fluid pressure and the energy density of the ideal fluid whereas $p_{(1)}$ and $e_{(1)}$ are the respective perturbations arising due to viscosity.
\begin{eqnarray}
e(r,\theta)= e_{(0)}(r,\theta)+\lambda  e_{(1)}(r,\theta) \nonumber \\
p(r,\theta)= p_{(0)}(r,\theta)+\lambda  p_{(1)}(r,\theta) 
\end{eqnarray}
 where $\lambda$ is the perturbation parameter. In the above series expansions, $e_{(1)}<<e_{(0)}$ and $p_{(1)}<<p_{(0)}$ so that the second and higher order perturbations in the energy density and pressure are small compared to zeroth order and first order terms and hence are discarded. 
 It is assumed here that the internal energy density is very small and the total energy is approximately equal to the rest-mass density i,e. $e \approx \rho$.
The normalization condition $u^{\alpha}u_{\alpha}=-1$ leads to,
\begin{equation}
-\left(1-\frac{2M}{r}\right)(u^{t})^2 + r^2 \sin^2 \theta (u^{\phi})^2=-1
\end{equation}
from which the temporal component of the fluid velocity is expressed as,
\begin{eqnarray}
u^{t}(r,\theta)=
\sqrt{\frac{1+ r^2 \sin ^2\theta (u^{\phi})^2}{\left(1-\frac{2 M}{r}\right)}}  \label{zero-lambda}
\end{eqnarray}
The angular velocity $\Omega(r,\theta)$ and the specific angular momentum $l(r,\theta)$ are related as, 
\begin{equation}
\Omega(r,\theta)=\displaystyle \frac{u^{\phi}(r,\theta)}{u^{t}(r,\theta)}, \qquad
l= -\frac{u_{\phi}(r,\theta)}{u_{t}(r,\theta)}, \qquad   \Omega(r,\theta)=\frac{\left(1-\frac{2M}{r}\right) }{r^2 \sin^2 \theta}l(r,\theta )
\end{equation} 
Substituting (\ref{zero-lambda}) in $\Omega(r,\theta)$ the azimuthal velocity $u^{\phi}$ can be easily expressed as follows,
\begin{equation}
u^{\phi}(r,\theta)= \displaystyle \frac{\Omega(r,\theta)}{\sqrt{(1-\frac{2 M}{r})-\Omega^2 r^2 \sin^2 \theta}}= \frac{(r-2M)^{1/2}\, l(r,\theta)}{r \sin ^2 \theta \left(r^3 + s^2 (2 M-r) \csc ^2\theta \right)^{1/2}}
\end{equation}
The general relativistic causal form of Navier-Stokes equation is,
\begin{eqnarray}
(e+p)a^{\mu}+\triangle^{\mu \rho}\nabla_{\rho}\,p +\pi^{\mu \alpha}a_{\alpha}
+ \triangle^{\mu}_{ \,\gamma} \triangle_{\kappa \tau}\nabla^{\tau}\pi^{\gamma \kappa} &= &0  
\label{Navier-2}
\end{eqnarray} 
where the four acceleration is $a^{\mu}=u^{\rho}\nabla_{\rho}u^{\mu}$ and the form of shear viscosity tensor is taken to be,
\begin{equation}
\pi^{\mu \nu}=-2\eta \sigma^{\mu \nu}+2 \tau_2 ^{<}D\eta \sigma^{\mu\nu}\,^{>}+\kappa_2 u_{\alpha} u_{\beta} R^{\alpha <\mu \nu >\beta}  \label{viscosity}
\end{equation} 
with following definitions,
\begin{eqnarray}
\sigma^{\mu \nu}&=&\triangle^{\mu \alpha}\triangle^{\nu \beta}\left(\displaystyle \frac{\nabla_{\alpha}u_{\beta}+\nabla_{\beta}u_{\alpha}}{2}\right)-\displaystyle \frac{1}{3} \triangle^{\mu \nu} \triangle^{\alpha \beta} \nabla_{\alpha}u_{\beta}\nonumber \\
^{<}D\sigma^{\mu\nu}\,^{>}&=&\triangle^{\mu \alpha}\triangle^{\nu \beta}\left(\displaystyle \frac{D\sigma_{\alpha \beta}+D\sigma_{\beta \alpha}}{2}\right)\displaystyle -\frac{1}{3} \triangle^{\mu \nu} \triangle^{\alpha \beta}D\sigma_{\alpha \beta}\nonumber \\
R^{\alpha <\mu \nu >\beta}&=&\triangle^{\mu \rho} \triangle^{\nu \sigma}\left(\displaystyle \frac{  R^{\alpha}_{\;\rho \sigma \gamma}g^{\beta \gamma}+ R^{\alpha}_{\; \sigma \rho \gamma}g^{\beta \gamma}}{2}\right) -\frac{1}{3} \triangle^{\mu \nu} \triangle^{\rho \sigma}  R^{\alpha}_{\;\rho \sigma \gamma}g^{\beta \gamma} 
\end{eqnarray}
Due to symmetry of the spacetime we have $D\eta= u^{\alpha}\nabla_{\alpha}\eta =0$ which implies $^{<}D(\eta \sigma^{\mu\nu})^{>}=\eta ^{<}D\sigma^{\mu\nu}\,^{>}$.
The constant specific angular momentum distribution $l(r,\theta)$ at the equatorial plane is described by $l(r,\theta) = s$ where $s$ is a constant parameter.
The first two terms of the momentum conservation equation (\ref{Navier-2}) correspond to the ideal fluid and (\ref{Navier-2}) reduces to the Euler equation when each of the coefficients individually vanish i,e. $\eta=\kappa_2=\tau_2=0$ implying $p_{(1)}=e_{(1)}=0$.
The contributions of third and forth terms of (\ref{Navier-2}) together with (\ref{viscosity}) are,
\begin{eqnarray}
\pi^{\mu \gamma}a_{\gamma}&=& \left[-2\eta \sigma^{\mu \gamma}+2 \tau_2 \eta ^{<}D \sigma^{\mu\gamma}\,^{>}+\kappa_2 u_{\alpha} u_{\beta} R^{\alpha <\mu \gamma>\beta}\right] a_{\gamma} \nonumber \\[1mm]
\text{and}\quad
\triangle^{\mu}_{ \,\gamma} \triangle_{\kappa \tau}\nabla^{\tau}\pi^{\gamma \kappa}&=&\triangle^{\mu}_{ \,\gamma} \triangle_{\kappa \tau}\nabla^{\tau}\left[-2\eta \sigma^{\gamma \kappa}+2 \tau_2 \eta ^{<}D \sigma^{\gamma \kappa}\,^{>}+\kappa_2 u_{\alpha} u_{\beta} R^{\alpha <\gamma \kappa >\beta}\right] \nonumber
\end{eqnarray} 
  for constant coefficients and constant value of specific angular momentum $l(r,\theta)=s$.
  Let us now consider the term $\pi^{\mu \gamma} a_{\gamma}$ for $\mu=t,r,\theta, \phi$ respectively.\\
For $\mu = t, \qquad \pi^{t \nu}a_{\nu} = \left[-2\eta \sigma^{t \nu}-\tau_2 ^{<}D(-2\eta \sigma^{t\nu})^{>}+\kappa_2 u_{\alpha} u_{\beta} R^{\alpha <t \nu >\beta}\right] a_{\nu} $  
\begin{flalign*}
&-2\eta \sigma^{t \nu} a_{\nu}= \frac{2\eta  s^2 r   \left[ M r (r-3M) \sin ^4 \theta -s^2 (1-2 M/r)^2 (r-3M\sin^2 \theta )\right]}{\sin \theta \sqrt{r-2M}\left(r^3 \sin^2 \theta + s^2 (2 M-r) \right)^{5/2}}, \nonumber \\[1pt]
&^{<}D(-2\eta \sigma^{t\nu})^{>}a_{\nu} =0,  \qquad  u_{\alpha} u_{\beta} R^{\alpha <t \nu >\beta}a_{\nu}=0 \nonumber 
\end{flalign*}
$\mu = r: \qquad \pi^{r \nu}a_{\nu} = \left[-2\eta \sigma^{r \nu}-\tau_2 ^{<}D(-2\eta \sigma^{r\nu})^{>}+\kappa_2 u_{\alpha} u_{\beta} R^{\alpha <r \nu >\beta}\right] a_{\nu}  $ 
\begin{eqnarray}
\sigma^{r \nu} a_{\nu}&=&0, \nonumber \\[1mm]
\tau_2  ^{<}D(-2\eta \sigma^{r\nu})^{>}a_{\nu} &=& \tau_2 \eta \frac{4s^2(r-3M) \left[M r(r-3 M) \sin^4 \theta - s^2   (1-2 M/r)^2 (r-3M \sin^2 \theta \right]}{
	\left(r^3 \sin^2 \theta + s^2  (2 M-r) \right)^3}, \nonumber \\[1mm]
	 \kappa_2 u_{\alpha} u_{\beta} R^{\alpha <r \nu >\beta}a_{\nu} &= & \frac{\kappa_2 M}{2r^5 \sin^2 \theta\left(r^3 \sin^2 \theta + s^2  (2 M-r) \right)^3} \left[s^2 (r-2M)^2 \sin^2 \theta \left\lbrace s^2  \left(8 M^2  s^2 \right. \right. \right. \nonumber \\
	\qquad && -\left. \left. \left. r^3(r-3M)(1-2 \sin^2 \theta)\right) -  2r^6(M+2) \sin^4 \theta \right. \right. \nonumber \\
	&&- \left. \left. r s^2  (r^3 + 3Mr^2 + 2rs^2-8 M s^2)\right\rbrace -  4M r^8  \right] \nonumber
\end{eqnarray} 
$\mu = \theta: \qquad \pi^{\theta \nu}a_{\nu} = \left[-2\eta \sigma^{\theta \nu}-\tau_2 ^{<}D(-2\eta \sigma^{\theta\nu})^{>}+\kappa_2 u_{\alpha} u_{\beta} R^{\alpha <\theta \nu >\beta}\right] a_{\nu} $ 
\begin{eqnarray}
\sigma^{\theta \nu} a_{\nu}&=&0, \nonumber \\[1mm]
\tau_2  ^{<}D(-2\eta \sigma^{\theta \nu})^{>}a_{\nu} &=&  \tau_2 \eta \frac{4 s^2 \cot \theta \left[ M r (r-3 M)\sin^4 \theta-s^2 (1-2 M/r)^2 (r-3M \sin^2 \theta)\right]}{ \left(r^3 \sin^2 \theta + s^2  (2 M-r)\right)^3},
\nonumber \\
\kappa_2 u_{\alpha} u_{\beta} R^{\alpha <\theta \nu >\beta}a_{\nu} &= & \kappa_2 \frac{  M s^2 \cot \theta (1-2M/r) \left[2(1-2M/r)-r^2 \sin^2 \theta \right]}{ r^3 \left(r^3 \sin^2 \theta + s^2  (2 M-r) \right)^2} \nonumber
\end{eqnarray}
$\mu = \phi: \qquad \pi^{\phi \nu}a_{\nu} = \left[-2\eta \sigma^{\phi \nu}-\tau_2 ^{<}D(-2\eta \sigma^{\phi\nu})^{>}+\kappa_2 u_{\alpha} u_{\beta} R^{\alpha <\phi \nu >\beta}\right] a_{\nu}  $ 
\begin{flalign*}
&-2\eta \sigma^{\phi \nu} a_{\nu}=\frac{2\eta  s r   \left[ M r (r-3M) \sin ^4 \theta -s^2 (1-2 M/r)^2 (r-3M\sin^2 \theta )\right]}{\sin \theta \sqrt{r-2M}\left(r^3 \sin^2 \theta + s^2 (2 M-r) \right)^{5/2}} = \frac{1}{s}(-2\eta \sigma^{t \nu} a_{\nu}), \nonumber \\[1pt]
&^{<}D(-2\eta \sigma^{\phi\nu})^{>}a_{\nu} =0, \quad
u_{\alpha} u_{\beta} R^{\alpha <\phi \nu >\beta}a_{\nu} =0  \nonumber
\end{flalign*} 
Thus for circular orbits of fluid flow the non-zero contributions of $\pi^{\mu \gamma} a_{\gamma}$ for $\mu=t,r,\theta, \phi$ are obtained to be,
\begin{eqnarray}
 \pi^{t \nu}a_{\nu}&=&-2 \eta \sigma^{t \nu}a_{\nu} \nonumber\\ 
\pi^{r \nu}a_{\nu}&=& \tau_2  ^{<}D(-2\eta \sigma^{r\nu})^{>}a_{\nu}+\kappa_2 u_{\alpha} u_{\beta} R^{\alpha <r \nu >\beta}a_{\nu} \nonumber \\
 \pi^{\theta \nu}a_{\nu}&=& \tau_2  ^{<}D(-2\eta \sigma^{\theta\nu})^{>}a_{\nu}+\kappa_2 u_{\alpha} u_{\beta} R^{\alpha <\theta \nu >\beta}a_{\nu} \nonumber \\ 
 \pi^{\phi \nu}a_{\nu}&=&-2 \eta \sigma^{\phi \nu}a_{\nu}=\frac{1}{s}(-2\eta \sigma^{t \nu} a_{\nu}) \label{third}
\end{eqnarray}
Let us now consider diferent components of the forth term in (\ref{Navier-2}).
\begin{eqnarray}
\mu = t: \qquad \triangle^{t}_{ \,\gamma} \triangle_{\kappa \tau}\nabla^{\tau} \left[-2\eta \sigma^{\gamma \kappa}-\tau_2 ^{<}D(-2\eta \sigma^{\gamma \kappa})^{>}+\kappa_2 u_{\alpha} u_{\beta} R^{\alpha <\gamma \kappa >\beta}\right]=0 \label{forth-1} \\[1mm]
\mu = \phi: \qquad \triangle^{\phi}_{ \,\gamma} \triangle_{\kappa \tau}\nabla^{\tau} \left[-2\eta \sigma^{\gamma \kappa}-\tau_2 ^{<}D(-2\eta \sigma^{\gamma \kappa})^{>}+\kappa_2 u_{\alpha} u_{\beta} R^{\alpha <\gamma \kappa >\beta}\right]=0 \label{forth-2}
\end{eqnarray}
\begin{eqnarray}
\mu = r: \qquad \triangle^{r}_{ \,\gamma} \triangle_{\kappa \tau}\nabla^{\tau} (-2\eta \sigma^{\gamma \kappa})&=&0 \nonumber \\ [2mm]
\triangle^{r}_{ \,\gamma} \triangle_{\kappa \tau}\nabla^{\tau}(-\tau_2 ^{<}D(-2\eta \sigma^{\gamma \kappa})^{>})&=& -\tau_2 \eta \frac{4  s^2 \sin^2 \theta (1-2M/r) (r-3 M) \left(2r^3+Ms^2-9Mr^2 \sin^2 \theta \right)}{\left(r^3 \sin^2 \theta +s^2 (2 M-r)\right)^3} \nonumber \\[1mm]
 \triangle^{r}_{ \,\gamma} \triangle_{\kappa \tau}\nabla^{\tau}[\kappa_2 u_{\alpha} u_{\beta} R^{\alpha <\gamma \kappa >\beta}]&=& -\kappa_2\frac{3 M s^2 \sin^2\theta (1-2M/r)(r-3M)}{r \left(r^3\sin^2\theta+s^2 (2 M-r) \right)^2} \nonumber
\end{eqnarray}
\begin{eqnarray}
 \mu = \theta :\qquad \triangle^{\theta}_{ \,\gamma} \triangle_{\kappa \tau}\nabla^{\tau} (-2\eta \sigma^{\gamma \kappa})&=&0 \nonumber \\ [2mm]
\triangle^{\theta}_{ \,\gamma} \triangle_{\kappa \tau}\nabla^{\tau}(-\tau_2 ^{<}D(-2\eta \sigma^{\gamma \kappa})^{>})&=& \tau_2 \eta \frac{4 \sin^2\theta (2M-r) s^2 \cot \theta \left(2 r^4-3M r^3 \sin^2 \theta - 5M r s^2+ 12M^2s^2\right)}{r^2 \left(r^3 \sin^2\theta+s^2(2M-r) \right)^3} \nonumber \\
\triangle^{\theta}_{ \,\gamma} \triangle_{\kappa \tau}\nabla^{\tau}[\kappa_2 u_{\alpha} u_{\beta} R^{\alpha <\gamma \kappa >\beta}]&=& \kappa_2\frac{3 \sin^2\theta (1-2M/r) s^2 \cot \theta }{r  \left(r^3\sin^2\theta+s^2 (2 M-r) \right)^2} \nonumber
\end{eqnarray} 
From above given computations, it is found that $-2\eta \sigma^{t \nu} a_{\nu}$ and $-2\eta \sigma^{\phi \nu} a_{\nu}$ contribute to temporal and azimuthal components of the Navier-Stokes equation. Then, using (\ref{third}), (\ref{forth-1}) and (\ref{forth-2}) lead to,
\begin{equation}
\frac{2\eta  s r   \left[ M r (r-3M) \sin ^4 \theta -s^2 (1-2 M/r)^2 (r-3M\sin^2 \theta )\right]}{\sin \theta \sqrt{r-2M}\left(r^3 \sin^2 \theta + s^2 (2 M-r) \right)^{5/2}}=0 \label{temp}
\end{equation}
Then at the equatorial plane, (\ref{temp}) reduces to,
\begin{equation}
(r-3M)\left[M r- s^2\left(1-\frac{2M}{r}\right)^2\right]=0
\end{equation}
consisting of the root $s=\displaystyle \frac{r\sqrt{M r}}{(r-2M)} \equiv l_{k}$ which is the Keplarian angular momentum.
To proceed, let the coefficients $\eta$, $\kappa_2$ act as perturbations in the system so that they are expressed as,
\begin{equation}
\eta= \lambda m_1,  \qquad \kappa_2=\lambda m_2  \label{transport}
\end{equation}
where $m_1$, $m_2$ are constant input parameters and eventually we put $\lambda=1$.
Substitution of the metric (\ref{metric-1}) together with (\ref{zero-lambda}) and (\ref{transport}) in (\ref{Navier-2}) gives rise to four components of Navier-Stokes equation in terms of two unknown variables namely $ p_{(1)}(r, \theta)$ and $e_{(1)}(r, \theta)$ and specific angular momentum distribution $l(r,\theta)$. For $l(r,\theta)=s$, the temporal and azimuthal components gives rise to (\ref{temp}) while the corrections terms $p_{(1)}$ and $e_{(1)}$ are determined from radial and angular components of (\ref{Navier-2}). 
Furthermore if there exists an equation of state (eos) relating $p_{(1)}$ and $e_{(1)}$ then the undetermined variable from the Navier-Stokes equation reduces to determining $p_{(1)}$. 
Let us suppose that the viscous fluid satisfies a barotropic equation of state. Expanding the equation of state up to linear order in $\lambda$, we obtain,
\begin{equation}
p_{(0)}+ \lambda p_{(1)}=K(e_{(0)}+\lambda e_{(1)})^{\gamma}
\end{equation} 
where $K$ is constant and $\gamma$ are constant polytropic exponent.
The equation of state at the zeroth order and the first order of $\lambda$ are respectively,
\begin{eqnarray}
p_{(0)}& =& K e_{(0)}^{\gamma} \qquad (\text{ideal fluid eos})\\[1mm]
p_{(1)}&=& \gamma K e_{(0)}^{\gamma-1}e_{(1)}  \label{e-corr} \label{correction-1}
\end{eqnarray}
The linear order energy density correction can then be determined by inverting (\ref{correction-1}) as follows,,  
\begin{eqnarray}
e_{(1)}=\frac{p_{(1)}}{\gamma K e_{(0)}^{\gamma-1}}
\end{eqnarray}
In the rest of discussion we will set the mass of the black hole $M=1$ and also put $K=1$. 
Then the radial component of (\ref{Navier-2}) in terms of $m_1$, $m_2$ and $\tau_2$ up to linear order in $\lambda$ is,
\begin{flalign}
&\frac{(\tau_2 m_1) s^2(r-3) }{2 r^2 \left(r^3\sin^2 \theta +s^2(2-r) \right)^3}
\left[r^3  \cos 4 \theta(10 r-21)  +\cos 2 \theta  \left\lbrace 4 r^3 (2r^2 -14r
+21)-8 s^2(r-3) (r-2)\right\rbrace  \right. \nonumber \\
&\left.\qquad \hspace{170pt} -\,r^3(2 r-7) (4 r-9) -8 s^2 (r-2) (r^2- 3r +3)\right]\nonumber \\[3pt]
&+ m_2 \frac{3 r^6+  r^6 \cos 4 \theta +2 r^3 \cos 2 \theta \left\lbrace s^2(r-2) (5 r-14)-2 r^3\right\rbrace -2 r^3 s^2(r-2) (5 r-14)
	-4s^4 (r-2)^3 }{4r^5 \left(r^3 \sin^2 \theta + s^2 (2-r) \right)^2}\nonumber \\[3pt]
&\,- \,	\frac{ \left\lbrace r^3 \sin^2 \theta - s^2 (2-r)^2 \right\rbrace \left(\gamma  K +  e_{(0)}^{1-\gamma }\right)}{\gamma  K r^2 \left(r^3 \sin^2 \theta + s^2 (2-r)\right) }p_{(1)} \,+\,\frac{  (r-2)}{r} \frac{\partial p_{(1)}}{\partial r} \quad =\quad 0
\label{NSE-radial}  
\end{flalign}
and similarly the angular part becomes,
\begin{flalign}
&\tau_2 m_1\frac{ 4 s^2 \cot \theta  \left\lbrace r^3(4 r-9) \sin^2 \theta +(r-2) \left(2 s^2 (4 r-9)-r s^2(r-2) \csc ^2 \theta -2r^4 \right)\right\rbrace }{r^2 \left(r^3 \sin^2 \theta +s^2(2-r) \right)^3} \nonumber\\
&+ m_2 \frac{2 s^2(r-2) \cot \theta  \left(2r^3 \sin^2 \theta +s^2(r-2)\right)}{r^3 \left(r^3 \sin^2 \theta +s^2 (2-r)\right)^2} 
- \frac{  (r-2) s^2 \cot \theta \left(\gamma  K +  e_{(0)}^{1-\gamma }\right)}{\gamma  K  \left(r^3 \sin^2 \theta + s^2(2-r) \right)} p_{(1)}+ \frac{\partial p_{(1)}}{\partial  \theta}\, =\,  0
\label{NSE-angular}
\end{flalign}
The lowest order of perturbation which corresponds to the no-viscosity scenario, the energy density and pressure of the fluid match to that of the ideal fluid and these quantities can be expressed in terms of the total potential $W(r,\theta)$ (in the Newtonian limit) and specific angular momentum distribution $l(r,\theta)$ as follows,
\begin{eqnarray}
e_{(0)}=\left[\frac{(\gamma -1)}{\gamma} \frac{(e^{W_{in}-W(r,\theta )}-1)}{  K}\right]^{\frac{1}{\gamma -1}}, \qquad
p_{(0)}= K\left[\frac{(\gamma -1)}{\gamma} \frac{(e^{W_{in}-W(r,\theta )}-1)}{  K}\right]^{\frac{\gamma}{\gamma -1}}
\end{eqnarray}  
with the boundary condition defined to be $W_{in} \rightarrow 0 $ as $r \rightarrow \infty $.
From (\ref{W-sch}), the equipotential for constant $l$ is given by $W(r,\theta)=W_{(Sch)}(r,\theta)$.
The linear correction to fluid pressure $p_{(1)}$ due to viscosity is obtained by eliminating $p_{(1)}$ from (\ref{NSE-angular}) and then substituting back in (\ref{NSE-radial}) which gives rise to a first order partial differential equation in $p_{(1)}$ as follows,
\begin{flalign}
& 2 (\tau_{2} m_{1}) \cot \theta \left[r^6(4 r-9)  \cos 6 \theta +2r^6 (4 r (3
r-11)+45)  + 32 s^4 (r-2)^2(r-3)
 (2r-3) \right.  \nonumber  \\[2pt]
&- \left.\cos 2 \theta  \left\lbrace r^6(4 r (8 r-31)+135) 
- 16  r^3 s^2 (r-2) (7 r^2-28 r+27) + 32 s^4
(r-2)^2 (3 r^2-11r +9) \right\rbrace \right.  \nonumber \\[2pt]
&+  \left. 2 r^3\cos 4 \theta  \left\lbrace r^3 (4r^2 - 20r+27)-2 s^2(r-2)
\left(6 r^2-26 r+27\right) \right\rbrace -4 r^3 s^2(r-2) \left(22 r^2-86 r+81\right)  \right]  \nonumber \\[2pt]
&+  \left(r^3 \cos 2 \theta -r^3+2 (r-2) s^2\right)^3\left[  2r (r-2)^2  \cot \theta   \frac{\partial p_{(1)}}{\partial r} 
- \frac{  \left(r^3 \cos 2 \theta -r^3+2 s^2
	(r-2)^2 \right)}{s^2}\frac{\partial p_{(1)}}{\partial \theta} \right]  \nonumber \\[2pt]
&+  3(m_2) (r-2) \cot \theta  \left[  r^6 \cos 6 \theta -6 r^3 \cos 4 \theta \left(r^3-s^2(r-2)^2
\right) 
+  \cos 2 \theta  \left\lbrace 15 r^6-24 r^3 s^2 (r-2)^2 
 \right.  \right.  \nonumber \\[2pt]
& +\left.  \left. 16 s^4(r-3) (r-2)^2 \right\rbrace 
 + 2
\left\lbrace 9r^3 s^2(r-2)^2 -8 s^4(r-3) (r-2)^2 -8 r s^6 (1-2/r)^4 -5 r^6\right\rbrace  \right]  \quad = \quad 0
\label{p-corr}
\end{flalign}
In terms of $g_{tt}$ the above equation becomes ,
\begin{flalign}
& 2 (\tau_{2} m_{1}) \cot \theta  \left [ r^2 \cos 6 \theta(4 g_{tt}+1) + 2r^2\left\lbrace 12r^2 g_{tt}^2-4 r g_{tt} +5\right\rbrace -32 s^4 g_{tt}^2\left(\frac{1}{r}+g_{tt}\right) \left(\frac{1}{r}-2g_{tt}\right) \right. \nonumber \\[2pt]
&- \left. \cos 2\theta \left\lbrace  r^2(32r^2 g_{tt}^2-4 r g_{tt}+15)+16s^2g_{tt}(7r^2 g_{tt}^2-1)+\frac{32}{r^2}s^4 g_{tt}^2(3r^2 g_{tt}^2 +r g_{tt}+1)\right\rbrace  \right.\nonumber \\[2pt]
&+ \left. 2 \cos 4 \theta \left\lbrace r^2(4r^2 g_{tt}^2+4 r g_{tt}+3)+2s^2g_{tt} (6r^2 g_{tt}^2+2 r g_{tt}-1)\right\rbrace 
+ 4 s^2 g_{tt} (22r^2 g_{tt}^2 -2 r g_{tt}-3) \right]\nonumber \\[2pt]
&+(r^2 \cos 2\theta -r^2-2g_{tt}s^2)^3 \left[2 \cot \theta \, g_{tt}^2 \frac{\partial p_{(1)}}{\partial r}-\frac{(r^2 \cos 2\theta -r^2-2r s^2g_{tt}^2)}{s^2}\frac{\partial p_{(1)}}{\partial \theta} \right]\nonumber \\[2pt]
&- 3 m_2g_{tt} \cot \theta \left[r^3 \cos 6 \theta -6r^2 \cos 4\theta \left(r-s^2 g_{tt}^2\right)+\cos 2 \theta \left\lbrace  15r^3-24r^2 s^2 g_{tt}^2-16s^4g_{tt}^2 (g_{tt}+1/r) \right\rbrace \right. \nonumber \\[2pt]
&+ \left. 2 \left\lbrace - 9s^2 r^2 g_{tt}+8s^2g_{tt}^2(g_{tt}+1/r)-8 g_{tt}^4 \frac{s^6}{r^2}-5r^3\right\rbrace \right] \quad = \quad 0
\end{flalign}
\subsection{Numerical results and impact of shear viscosity on the shape of the disk}
The pressure correction equation (\ref{p-corr}) is numerically solved for $p_{(1)}$ for constant specific angular momentum and under the consideration that at a sufficiently large distance away from the horizon of the Schwarzschild black hole, the effects of viscosity are vanishingly small and hence can be neglected. 
The coefficients $\eta$, $\kappa_2$ and $\tau_2$ may in principle be functions of fluid presure but for simplicity are taken constants in our analysis. Here $\tau_2$ behaves as the relaxation time coefficient in the same spirit of IS formalism preserving causality of the relativistic Navier-Stokes equation and in the present case does not possess perturbative character. \\  
Mostly relevant for study concerning astrophysical black holes, the value of the ploytropic index is taken to be $\gamma=5/3$ with $K=1$ and $\tau_2=0.2$.
The effects of viscosity and curvature on the thick accretion disk are then estimated by comparing the profiles of constant pressure surfaces of the perfect fluid and that of the constant pressure surfaces arising due to viscous fluid characterised by small values of $\eta (=m_1)$ and $\kappa_2(=m_2)$.
A comparison of pressure plots between the perfect fluid given by $p_{(0)}$ and that of viscous fluid specified by $p_{(0)}+p_{(1)}$ for two different values of specific angular momentum namely $l_{ms}<s=3.8<l_{mb}$ and $s=l_{mb}=4$ with $\triangle W_{in}>0$ are depicted in FIG.1 and FIG.2 where $l_{ms}=3.67$ and $l_{mb}=4$ are the values of specific angular momentum corresponding to last stable (marginally) orbit and last bound orbit respectively of the Schwarschild black hole. 
In every set of the plot for a given value of $s$, we concentrate on same set of fluid profiles to determine the changes induced as a consequence of introducing viscosity and curvature through the correction $p_{(1)}$. It can be seen that the for $s=3.8$, cusps appear as a result of viscosity. 
\begin{figure}[H]
	
	\includegraphics[width=0.068 \hsize]{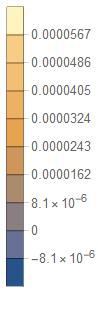}
	\includegraphics[width=0.23\hsize]{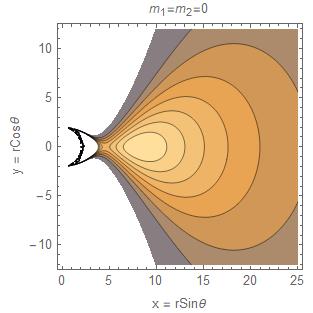} %
	\includegraphics[width=0.23\hsize]{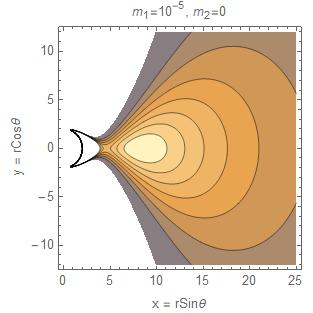}%
	\includegraphics[width=0.23\hsize]{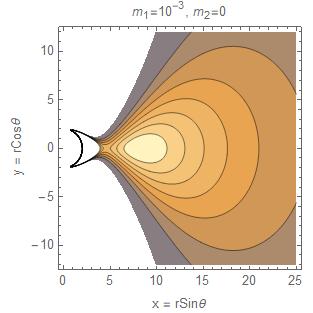}%
	\includegraphics[width=0.23\hsize]{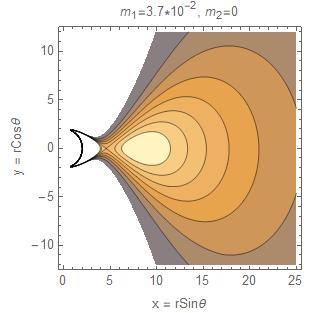}
	\\[1mm]
	\includegraphics[width=0.068 \hsize]{l=3_8-scale.jpg}
	\includegraphics[width=0.23\hsize]{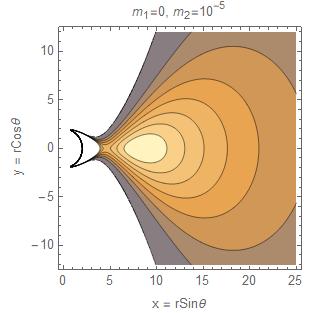} %
	\includegraphics[width=0.23\hsize]{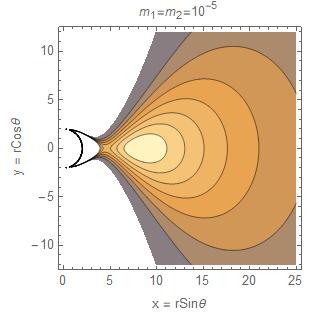}%
	\includegraphics[width=0.23\hsize]{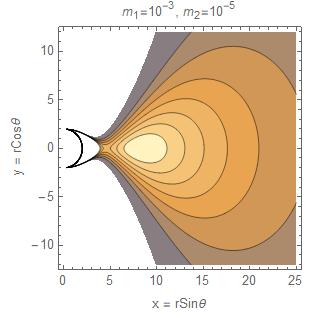}%
	\includegraphics[width=0.23\hsize]{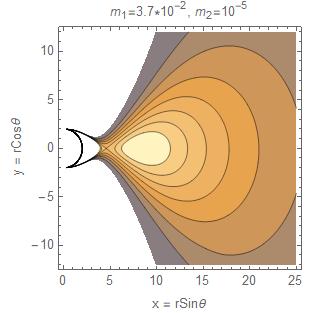}
	\\[1mm]
	\includegraphics[width=0.068 \hsize]{l=3_8-scale.jpg}
	\includegraphics[width=0.23\hsize]{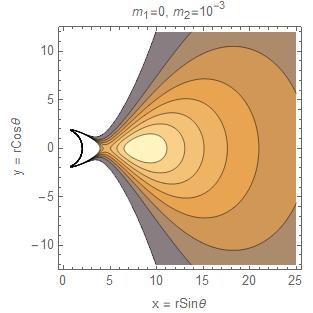} %
	\includegraphics[width=0.23\hsize]{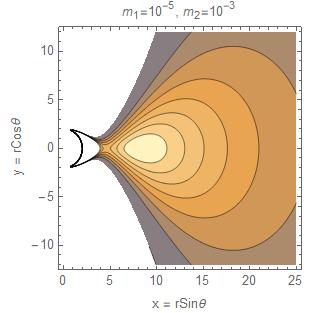}%
	\includegraphics[width=0.23\hsize]{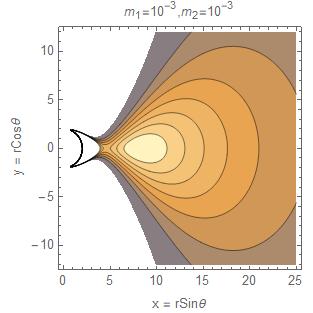}%
	\includegraphics[width=0.23\hsize]{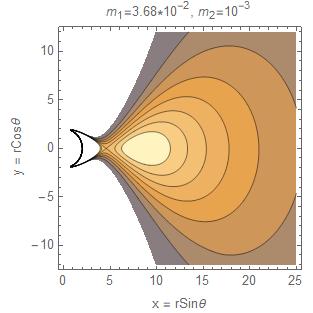}
	\\[1mm]
	\includegraphics[width=0.068 \hsize]{l=3_8-scale.jpg}
	\includegraphics[width=0.23\hsize]{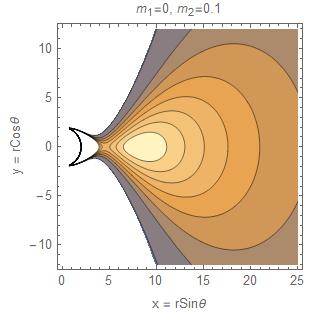} %
	\includegraphics[width=0.23\hsize]{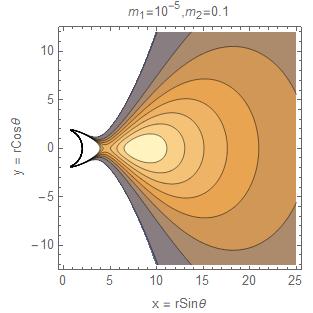}%
	\includegraphics[width=0.23\hsize]{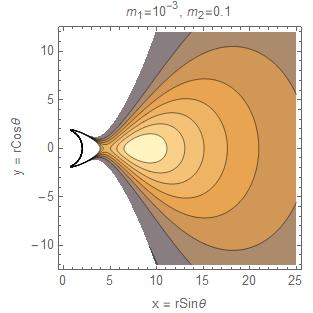}%
	\includegraphics[width=0.23\hsize]{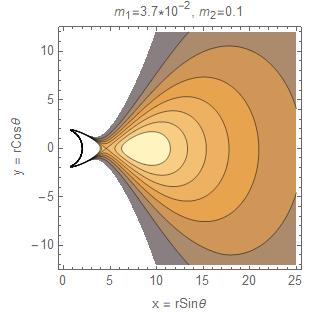}
	\caption{Comparison of constant pressure surfaces of ideal fluid ($m_1 = m_2 = 0$) and viscous fluid for $l_{ms}<s=3.8<l_{mb}$.}
\end{figure}
\begin{figure}[H]
	\includegraphics[width=0.068 \hsize]{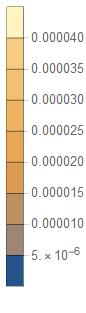}
	\includegraphics[width=0.23\hsize]{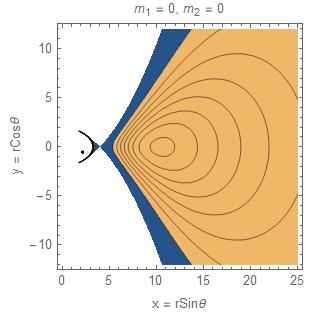} %
	\includegraphics[width=0.23\hsize]{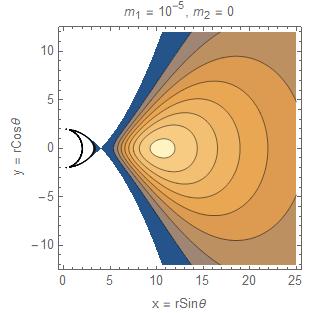}%
	\includegraphics[width=0.23\hsize]{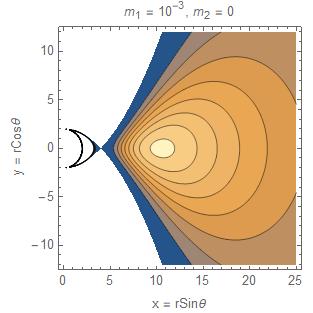}%
	\includegraphics[width=0.23\hsize]{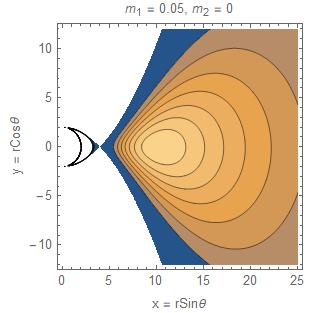}
	\\[1mm]
	\includegraphics[width=0.068 \hsize]{l=4-scale.jpg}
	\includegraphics[width=0.23\hsize]{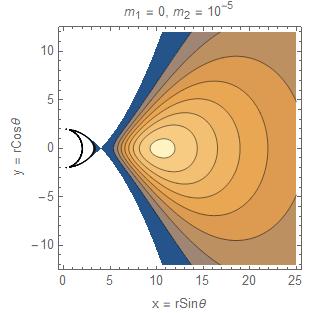} %
	\includegraphics[width=0.23\hsize]{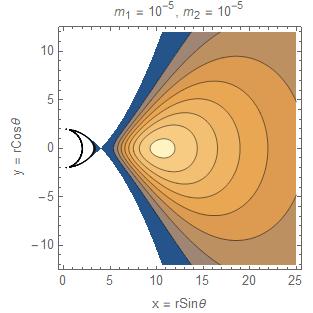}%
	\includegraphics[width=0.23\hsize]{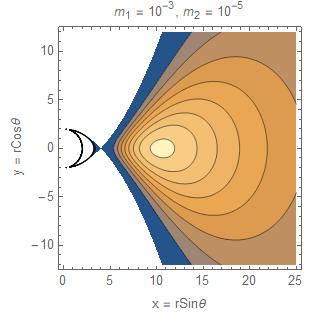}%
	\includegraphics[width=0.23\hsize]{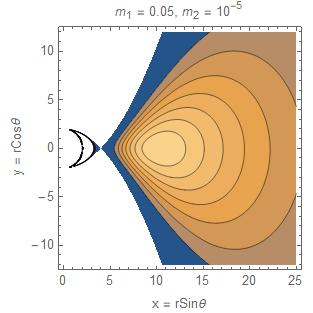}%
	\\[1mm]
	\includegraphics[width=0.068 \hsize]{l=4-scale.jpg}
	\includegraphics[width=0.23\hsize]{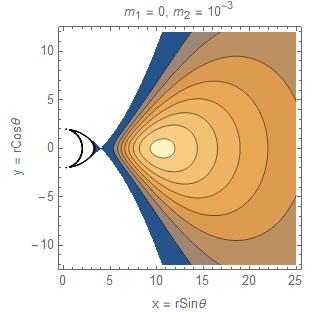} %
	\includegraphics[width=0.23\hsize]{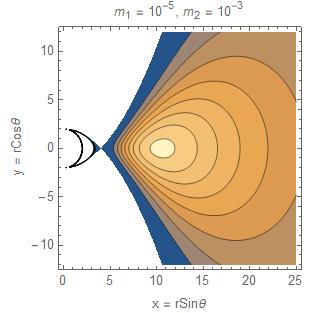}%
	\includegraphics[width=0.23\hsize]{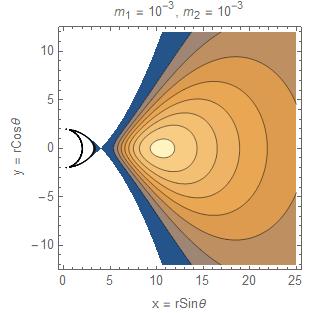}%
	\includegraphics[width=0.23\hsize]{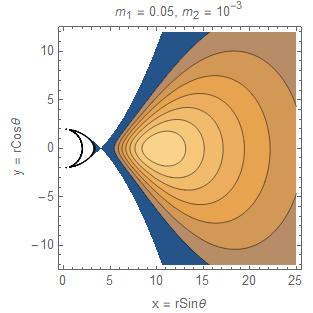}%
	\\[1mm]
	\includegraphics[width=0.068 \hsize]{l=4-scale.jpg}
	\includegraphics[width=0.23\hsize]{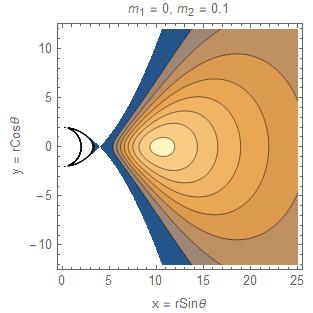} %
	\includegraphics[width=0.23\hsize]{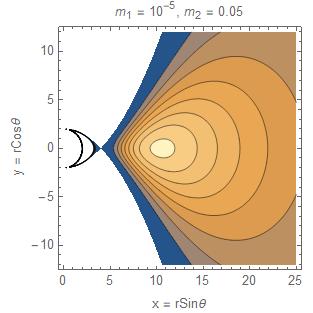}%
	\includegraphics[width=0.23\hsize]{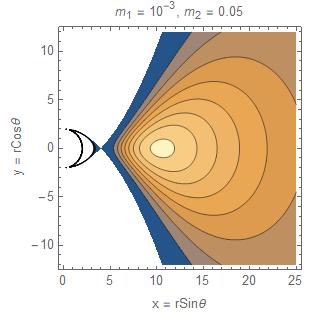}%
	\includegraphics[width=0.23\hsize]{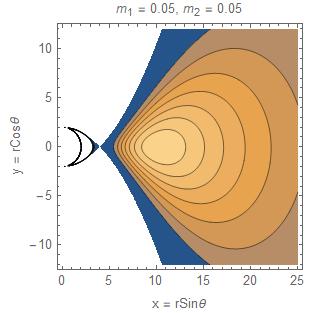}%
	\caption{Comparison of constant pressure surfaces of ideal fluid ($m_1 = m_2 = 0$) and viscous fluid for $s=4$.}
\end{figure}
\subsection{Integrability condition of general relativistic causal Navier-Stokes equation}
In order to address the integrability condition of the relativistic Navier-Stokes equation for a viscous fluid, we will follow the similar procedure which is adopted for studying the existence of the condition with an ideal fluid. 
The symmetry of the spacetime governs all flow parameters including the fluid pressure and hence the energy density to be functions of $r$ and $\theta$ only, on the other hand, the coefficients $\eta$, $\kappa_2$ and $\tau_2$ appearing in the shear viscosity tensor may in principle be functions of fluid pressure and further related to energy density.
Hence in this generalized scenario, the coefficients become  $\eta \equiv \eta(p), \kappa_2 \equiv \kappa_2(p)$ and $\tau_{2} \equiv \tau_{2}(p)$ and become eventually functions of radial and angular co-ordinates. 
So, the integrability condition for the general relativistic causal Navier-Stokes equation (\ref{Navier-2}) may be determined by fulfilling the following condition,
\begin{equation}
 \partial_{r}\partial_{\theta}p=\partial_{\theta} \partial_{r}p \label{cond}
\end{equation}    
which is also known as compatibility condition similar to the ideal fluid case. \\
Additionally due to the symmetry of the spacetime we find that $\triangle^{\mu \rho} \nabla_{\rho}p = \partial^{\mu} p$ so that the radial and angular components of (\ref{Navier-2}) may be written as,
\begin{eqnarray}
\partial_r p &=& (e+p)a_{r}+\pi_{r \alpha}a^{\alpha} +\triangle_{r \gamma} \triangle_{\kappa}^{\tau}\nabla_{\tau} \pi^{\gamma \kappa} = R(r,\theta)  \label{inte-1}\\[2mm]
\partial_{\theta} p &=& (e+p)a_{\theta}+\pi_{\theta \alpha}a^{\alpha} +\triangle_{\theta \gamma} \triangle_{\kappa}^{\tau}\nabla_{\tau} \pi^{\gamma \kappa}= Q(r,\theta) \label{inte-2}
\end{eqnarray}
 where  $\pi^{\mu \nu}=\left[-2\eta(p) \sigma^{\mu \nu}-\tau_2(p) ^{<}D(-2\eta(p) \sigma^{\mu\nu})^{>}+\kappa_2(p) u_{\alpha} u_{\beta} R^{\alpha <\mu \nu >\beta}\right] $  is the viscosity tensor and the four-acceleration is given by $a_{i} = \left[\partial_{i} \ln |u_{t}|-\displaystyle \frac{\Omega}{1-\Omega l}\,\partial_{i} l \right]$ with $i=r,\theta$. 
 Here $p$ and $e$ depict the total fluid pressure and the total energy density involving the ideal fluid part and the viscous corrections. 
 So (\ref{cond}) suggests that,
\begin{equation}
\partial_{\theta}R(r,\theta) = \partial_{r}Q(r,\theta) \label{final-inte}
\end{equation}
\begin{equation}
\Rightarrow \partial_{\theta} \left\lbrace (e +p)a_{r}\right\rbrace + \partial_{\theta} \left\lbrace \pi_{r \alpha}a^{\alpha}+\triangle_{r \gamma} \triangle_{\kappa}^{\tau} \nabla_{\tau} \pi^{\gamma \kappa} \right\rbrace = \partial_{r} \left\lbrace (e +p)a_{\theta}\right\rbrace + \partial_{r} \left\lbrace \pi_{\theta \alpha}a^{\alpha}+\triangle_{\theta \gamma} \triangle_{\kappa}^{\tau} \nabla_{\tau} \pi^{\gamma \kappa} \right\rbrace  \label{1} 
\end{equation}
As a result (\ref{1}) reduces to,
\begin{eqnarray}
\partial_{\theta} R(r,\theta) &=&
\partial_{\theta} \left\lbrace (e +p)a_{r}\right\rbrace +\left[-2   \partial_{\theta} \left\lbrace \eta(p)\sigma_{r \alpha} \right\rbrace -2\partial_{\theta}\left\lbrace \tau_2(p) ^{<}D(\eta(p)g_{r \mu} \,g_{\alpha \nu} \sigma^{\mu\nu})^{>} \right\rbrace \right.\nonumber \\[2mm] 
&&+ \left. \partial_{\theta}\left\lbrace \kappa_2(p) u_{\alpha} u_{\beta} g_{r \mu} g_{\alpha \nu}R^{\alpha <\mu \nu >\beta}\right\rbrace \right ] a^{\alpha} +\pi_{r \alpha} \partial_{\theta}a^{\alpha}  \label{2}\\[2mm]
\text{and} \qquad 
\partial_{r} Q(r,\theta) &=&
\partial_{r} \left\lbrace (e +p)a_{\theta}\right\rbrace +\left[-2   \partial_{r} \left\lbrace \eta(p)\sigma_{\theta \alpha} \right\rbrace -2\partial_{r}\left\lbrace \tau_2(p) ^{<}D(\eta(p)g_{\theta \mu}\, g_{\alpha \nu} \sigma^{\mu\nu})^{>} \right\rbrace \right.\nonumber \\[2mm] 
&&+ \left. \partial_{r}\left\lbrace \kappa_2(p) u_{\alpha} u_{\beta} g_{\theta \mu} g_{\alpha \nu}R^{\alpha <\mu \nu >\beta}\right\rbrace \right ] a^{\alpha} +\pi_{\theta \alpha} \partial_{r}a^{\alpha} \label{3}
\end{eqnarray}
Since we have obtained the solution of the fluid pressure correction by taking the coefficients $\eta, \tau_2$ and $\kappa_2$  as constants independent of functional dependences on radial and angular co-ordinates with constant angular momentum distribution, we stick to same assumptions for checking the intergrability condition in the viscous case. 
Therefore in order to determine the validity of integrability condition, we plot the $\partial_{r} \partial_{\theta}p=\partial_{\theta} \partial_{r}p$ togather with (\ref{2}) and (\ref{3}) for the value of constant angular momemtum $s=3.8$ for two different values of $\eta$ and $\kappa_2$. 
The validity of (\ref{cond}) is independent of choice of $\triangle W_{in}$.
\begin{figure}[h]
	\includegraphics[width=0.4 \hsize]{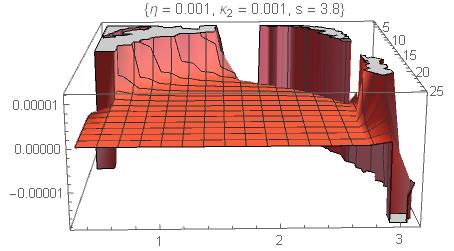}
	\includegraphics[width=0.4 \hsize]{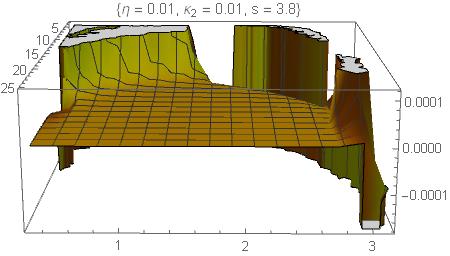}
	\caption{}
\end{figure}

 \section{Discussions and Outlook}
 The present work is devoted to understanding the impact of shear viscosity on the structure of relativistic geometrically thick accretion disks in the Schwarzschild geometry by investigating stationary solutions of the general relativistic causal Navier-Stokes equation. 
 As the present study involves first ever step towards comprehending the role of viscosity in a thick disk, we concentrate ourselves on obtaining stationary solutions of the fluid flow undergoing circular orbits around the Schwarzschild black hole in order to be able to compare our results with ideal fluid thick disks. So any fluid velocity perturbations as well as radial component of fluid four-velocity have been discarded. 
 Consequently due to the symmetry of the spacetime all flow parameters have remained time independent in our analysis.  
 The advantages of the approach adopted in the present work are (a) the causal theory of relativistic
 viscous hydrodynamics provides a scope for studying relativistic theory of stationary solutions of a viscous thick disk, (b) presence of additional curvature term in the shear viscosity tensor allows us to examine how curvature effects of the central black hole might have any influence on the shape of the thick disk. 
 Thus in contrary to traditional approaches, the curvature of the black hole, in our approach, has a direct influence on the motion of the fluid in a thick disk giving rise to different shapes of a thick disk in different external geometries of a central black hole.
 As an initial step, we have taken the Schwarschild black hole in the present study. \\  
 As a starting point for investigating structure of the thick disk in presence of viscosity, we have completely ignored any emergence of turbulence in the fluid and adopted several simplifying assumptions to capture the essence of viscosity by visualizing its effects as corrections to the ideal fluid configuration of the disk.
 The idea is that viscosity instils perturbative effects in the fluid pressure and hence in the energy density without giving rise to radial velocity, other velocity perturbations and heat flow in the fluid. 
 Thus we assume that the circular orbits are still allowed in this scenario and the causal Navier-Stokes equation are studied up to leading order in perturbation. 
 The zeroth order in perturbation corresponds to the disk described by the ideal fluid, whereas the shear viscosity with all constant coefficients in our case constitutes the linear order of perturbation.  
 To incorporate the causal prescription suggested in IS formalism, the term  $2 \tau_2 ^{<} D \eta \sigma^{\mu \nu} \,^{>}$ in the relativistic Navier-Stokes equation is considered in addition to the curvature terms so as to probe their impacts on the structure of the thick disk.
 Moreover the bulk viscosity does not play any role in our study.
By employing a barotropic equation of state for the viscous fluid, 
 the causal Navier-Stokes equation is numerically solved to determine the presure corrections $p_{(1)}$ under the assumption that $\tau_2$ is substantially large compared to $\eta,\kappa_2$, while both $\eta,\kappa_2$ act as perturbations. 
 We note here that $p_{(1)}$ vanishes if each of the coefficients individually vanish and in that case the disk is characterised by an ideal fluid.
 The influence of viscosity and the curvature on a thick disk is investigated by comparing the constant pressure
 surfaces characterized by ideal fluid pressure $p_{(0)}$ with same set of constant pressure surfaces described by  $p_{(0)}+p_{(1)}$ for $s=3.8$ and $s=4$ with constant values of $\eta$ and $\kappa_2$ and $\tau_2$. 
 For the value of specific angular momentum $l_{ms}<s=3.8<l_{mb}$, we showed that in comparison to the ideal fluid, in presence of viscosity and curvature, one can obtain cusps or self intersection of the pressure surfaces for which the constant pressure surfaces of the ideal fluid do not self-intersect. 
 The formation of cusps of constant pressure surfaces at locations different from those predicted with the ideal fluid is a direct consequence of viscosity and curvature of the Schwarschild spacetime. 
 This behaviour can be seen in figure 1 where the comparison has been performed with different values of $\eta$ and $\kappa_2$. 
 An important aspect of our study is that the causal prescription of the Navier-Stokes equation is an ubiquitous requirement for constructing the viscous relativistic thick disk. On the other hand as shown in the Fig. 1, it is  possible to construct the relativistic disk with the help of curvature terms which would behave as shear viscosity in the fliud. 
 We worked with $s=l_{mb}=4$ and observed disappearence of constant pressure surfaces for large values of $\eta \approx 0.05$ independent of the values of $\kappa_2$.\\
 The validity of the integrability condition is checked in the present work by assuming all coefficients namely $\eta$, $\tau_2$, and $\kappa_2$ as constants. It can be seen from (\ref{inte-1}) and (\ref{inte-2}) that without the causal prescription and under the assumption of circular orbits, the shear viscosity term $-2\eta \sigma^{\mu \nu}$ does not alter the integrabilty condition (\ref{final-inte}) resulting into vanishing of pressure and energy density corrections. This is also corroborated by the pressure plots in figure 1 and figure 2 with $s=3.8$ and $s=4.0$ where it can be observed that the fluid pressure correction $p_{(1)}$ cannot be expressed only in terms of $\eta$ independently of relaxation time coefficient $\tau_2$ hence the causal prescription is essential for the construction of a disk with circular fluid flow in addition to curvature effects. 
 The integrability condition of general relativistic Navier-Stokes eqution is checked for different values of $\eta$ and $\kappa_2$ with constant specific angular momentum distribution and the violation occurs of the order of $\eta \kappa_2$ i,e. of the order of second order in perturbation. This is acceptable in present context as we obtain solutions by working up to linear order of perturbation. In the following, we briefly mention immediate extentions
 of the present work which includes study of the shapes of the viscous thick disk with non-constant angular momentum distributions. 
 In presence of voscosity, this work can also be extended in the rotating background, for example in the context of Kerr black hole, for studying stationary solutions with constant/non-constant specific angular momentum distributions. The obtained results can then be utilized for examining cumulative effects of viscosity, curvature and spin of the Kerr black hole on the shape of the disk.
 Since the present work is the first step towards understanding the role of viscosity in a thick disk, we have imposed several simplified assuptions. 
 A more generalized scenario, worthwhile to study, would be to include heat flow, velocity perturbations and a non-zero radial velocity giving rise to perturbations in pressure and energy densities. 
 We currently leave these issues for future investigations. 

\begin{acknowledgments}
	We gratefully thank Rudolf Baier, Paul Romatschke, Jose. A. Font and Isabel C. Carrion for fruitful and illuminating discussions during various stages of this work. We are also thankful to Volker Perlick for helpful suggestions. We express our gratitude to Cluster of Excellence
	``Quantum Frontiers" and the Research training Group 1620 ``Models of Gravity" funded by Deutsche Forschungsgemeinschaft (DFG).
	This work is supported by DFG with grant/40401089.
\end{acknowledgments}

\end{document}